# Imaging Polarimetry and Photometry of Comet 21P/Giacobini-Zinner

Ekaterina Chornaya[1,2,*], Evgenij Zubko[1], Igor Luk'yanyk[3], Anton Kochergin[1,2], Maxim Zheltobryukhov[2], Oleksandra V. Ivanova[4,3,5], Gennady Kornienko[2], Alexey Matkin[2], Alexander Baransky[3], and Gorden Videen[6,7]

[1] Far Eastern Federal University, 8 Sukhanova St., Vladivostok 690950, Russia (rainu13rainu@gmail.com);

[2] Ussuriysk Astrophysical Observatory of FEB RAS, 21 Solnechnaya St., Gornotaezhnoe 692533, Russia;

[3] Astronomical Observatory, Taras Shevchenko National University of Kyiv, 3 Observatorna St., 04053, Kyiv, Ukraine

[4] Astronomical Institute of the Slovak Academy of Sciences, SK-05960 Tatranská Lomnica, Slovak Republic

[5] Main Astronomical Observatory of National Academy of Sciences, Kyiv, Ukraine

[6] Space Science Institute, 4750 Walnut Street, Boulder Suite 205, CO 80301, USA

[7] Department of Astronomy and Space Science, Kyung Hee University, 1732, Deogyeong-daero, Giheung-gu, Yongin-si, Gyeonggi-do 17104, South Korea

**Abstract.**

We report results of our polarimetric observations of comet 21P/Giacobini-Zinner made at phase angles, α ≈ 76–78°, between 10 and 17 of September, 2018, and compare them with previous measurements. We find significant variations in the polarimetric signals that appear consistent with those reported previously. These variations and subsequent modeling suggest that the particles in the coma are replenished within a period of approximately one day. This period is significantly shorter for highly absorbing carbonaceous particles than for non-absorbing Mg-rich silicate particles. Such a difference in the relative abundances of these components can lead to variations in the polarization response of the coma. The strong positive polarization in the subsolar direction suggests a large relative abundance of carbonaceous material, which may be an indicator of jet-type activity.

\* Corresponding Author

E-mail address: rainu13rainu@gmail.com

Phone: +7 999 040 03 45 (UTC+10)

Fax: +7 4234 39 11 21

**Highlights**

We observe degree of linear polarization and color in comet 21P/Giacobini-Zinner.

Comet reveals noticeable spatial and temporal variations of polarization.

Observations can be reproduced under assumption of two-component composition of coma.

The 21P coma consists of Mg-rich silicate and amorphous-carbon particles.

Relative abundance of these two components is inferred on different epochs.

# 1. Introduction

Comet 21P/Giacobini–Zinner (hereafter 21P/G-Z) is a short-period comet revolving around the Sun with period of approximately 6.6 year along an elongated orbit. Its perihelion lies at about 1 au making ground-based observations at large phase angles possible. This comet has a long history of investigation, including polarimetric observations, which, were accomplished as early as 1959 (Martel 1960). In addition, the polarimetry of comet 21P/G-Z has been observed in its apparitions of 1985 (Chernova et al 1993; Kurchakov et al. 1986) and 1998–1999 (Kiselev et al. 2000). In this paper we report results of a polarimetric survey of the 2018 apparition.

Numerous comets reveal *red polarimetric color* (e.g., Chernova et al 1993; Levasseur-Regourd et al. 1996), implying that the degree of linear polarization $P$ at a longer wavelength $\lambda$ exceeds $P$ at a shorter wavelength; however, comet 21P/G-Z seemingly does not obey this trend. Kiselev et al. (2000) found the polarization of this comet in the blue continuum filter to be somewhat higher than in the red continuum filter in the 1998–1999 apparition, i.e. *blue polarimetric color*. To support their finding, Kiselev et al. (2000) referred to the polarimetric measurements of 21P/G-Z by Kurchakov et al. (1986) that unfortunately were published only in Russian and, therefore, did not gather significant attention.

The findings of blue polarimetric color by Kurchakov et al. (1986) are somewhat ambiguous. First of all, the uncertainty of the results obtained with the *International Halley Watch* (*IHW*) blue continuum (*BC*) filter and red continuum (*RC*) filter ($\lambda$ = 0.4845 µm and 0.684 µm, respectively) have significant uncertainties associated with them. The individual polarizations of the measurement obtained on 1985-09-16 were $P_{BC}$ = (18.2 ± 3.2)% and $P_{RC}$ = (16.9 ± 1.6)%. The uncertainties of each of these measurements is greater than the difference between the polarizations measured at the different colors, and the uncertainty determined using the blue filter is more than twice the difference. Second, Kurchakov et al. (1986) also measured the linear polarization with the wideband $V$ filter (five epochs in September of 1985) and with the wideband $B$ and $R$ filters (six epochs for each). On one of these epochs, there was no clear wavelength dependence of $P$ larger than the error bars, i.e., *neutral polarimetric color*, but on all the additional epochs, the comet displayed *red polarimetric color*. It is worth noting that 21P/G-Z is a very dusty comet. In the literature, the relative contribution of dust and gases into the light-scattering response is quantified with the parameter $W_{4845}$ that is the ratio of fluxes measured through the *IHW C2* filter ($\lambda$ = 0.514 µm) over that with the *IHW BC* filter; this quantity is measured in Å (e.g., Krishna Swamy 1986). In comet 21P/G-Z $W_{4845}$ = 50 Å (Chernova et al. 1993) that corresponds to very little relative contribution of the gaseous emission, ~1% (Zubko et al. 2016). Furthermore, a weak contribution of gaseous emission could be deduced from the weak difference between the polarimetric response measured with the wideband $B$ and $R$ filters and with the narrowband continuum filters *IHW BC* and *RC*. This difference in Kurchakov et al. (1986) does not exceed the error bars. Therefore, the polarimetric response of comet 21P/G-Z in the wideband filters corresponds to its dust.

Although the findings of Kurchakov et al. (1986) differ from those reported by Kiselev et al. (2000), this does not necessarily suggest that some of those observations are inconsistent. These two teams observed the comet in different apparitions and also using different apertures that could explain their divergence. While blue polarimetric color is uncommon, it is not unreported. Blue polarimetric color was detected *in situ* in the innermost coma of 1P/Halley using the *Giotto* spacecraft (Levasseur-Regourd et al. 2005). The blue polarimetric color reported by Kiselev et al. (2000) does not appear to be a persistent feature of comet 21P/G-Z, since most of the previous measurements were of red polarimetric color. It seems reasonable to suggest that the polarimetric color is subject to temporal variations in 21P/G-Z. Such a conclusion is in accordance with variations of the photometric color that was found in some comets (i.e., Ivanova et al. 2017; Luk'yanyk et al. 2019). The present study is aimed at investigating the temporal variations of the polarization of comet 21P/G-Z that may help better understand the seeming inconsistencies in the previous measurements.

## 2. Observations and data reductions

*2.1. Imaging polarimetry*

In September 2018, we conducted polarimetric observations of comet 21P/G-Z. Weather conditions were favorable on four dates, September 10, 12, 16, and 17. Observations were made at the prime focus of the 22-cm telescope of the Ussuriysk Astrophysical Observatory (observatory code C15), which operates within the International Scientific Optical Network (ISON). The telescope was equipped with a commercially available CCD detector SBIG STX-16803 that has a resolution of 4096 x 4096 pixels and pixel size of 9 μm. The field of view of the CCD detector is 251 x 251 arcmin with angular resolution of 3.68 x 3.68 arcsec per pixel. The comet was observed through the *V* filter of the Johnson-Cousins photometric system. In addition, we use a dichroic polarization filter (analyzer). The analyzer was rotated sequentially through three fixed position angles 0°, +60°, and +120°. Log of observations is given in Table 1.

The obtained images of the comet have been processed using the Image Reduction and Analysis Facility (IRAF) software system for reduction and analysis of astronomical data. It includes basic programs for bias subtraction, removal of cosmic-ray events and flat-field correction. Flat-field correction was accomplished with images of the twilight sky. For more accurate data analysis, we use a procedure that retrieves histogram of counts of the sky background. This allows us to determine accurately its maximum level and then subtract it from every image. For calibration purposes, we observed polarized and non-polarized standard stars from the lists of Hsu & Breger (1982), Schmidt et al. (1992), and Heiles (2000). It revealed the instrumental polarization being about 0.3 percent at good atmospheric conditions.

The resulting images were used to infer polarimetric images of the comet and compute the degree of linear polarization at given apertures using the Fesenkov formulae (e.g., Fessenkoff 1935; Hines et al. 2014). Within this approach, the Stokes parameters corresponding to the total intensity (*I*) and to the linear polarization (*Q* and *U*) are defined as follows:

$$I = \frac{2}{3}(F_0 + F_{60} + F_{120}),$$

$$Q = \frac{2}{3}(2F_0 - F_{60} - F_{120}), \qquad (1)$$

$$U = \frac{2}{\sqrt{3}}(F_{60} - F_{120}).$$

where, $F_0$, $F_{60}$, and $F_{120}$ stand for fluxes of electromagnetic radiation from the comet measured with different orientations of the analyzer. The total degree of linear polarization *P* is defined then:

$$P = \frac{\sqrt{Q^2 + U^2}}{I} \cdot 100\%. \qquad (2)$$

Linear polarization is further characterized by *position angle* θ of the *normal* to its plane. This angle is determined by the Stokes parameter $Q$ and $U$:

$$\theta = \frac{1}{2}\text{atan}\left(\frac{U}{Q}\right) + \theta_0, \qquad (3)$$

where, $\theta_0 = 180°$ if $Q > 0$ and $U < 0$, $\theta_0 = 0°$ if $Q > 0$ and $U \geq 0$, or $\theta_0 = 90°$ if $Q < 0$ (e.g., Zubko et al. 2015).

It is important to note that in comets, orientation angle θ takes on one out of two values, either 0° (180°) or 90°, with only slight perturbations from those two stable orientations (e.g., Zubko et al. 2015). It means that on average the plane of linear polarization is oriented either perpendicular to the scattering plane (defined by the Sun, comet, and observer) or it lies within the scattering plane. In practice, this implies that dust particles in a comet appear predominantly in random orientations (Bohren and Huffman 1983), although there are presumably some exceptions from this general trend in a limited part of the coma (Moreno et al. 2018). As one can see in Table 2, our observations of comet 21P/G-Z reveal only slight deviations of θ from 0°, ≤ 15.2°, that is very much typical for other comets (Zubko et al. 2015).

In an optically thin cloud of randomly oriented dust particles, that appears to be a good model of cometary coma, the degree of linear polarization $P$ given in Eqs. (2) and (3) can be reduced to $P_r$:

$$P_r = P\cos(2\theta) = -\frac{Q}{I} \cdot 100\% = \frac{F_\perp - F_\parallel}{F_\perp + F_\parallel} \cdot 100\%. \qquad (4)$$

Here, $F_\perp$ and $F_\parallel$ stand for fluxes of light polarized perpendicular to the scattering plane and within the scattering plane, respectively. Unlike $P$, the degree of linear polarization $P_r$ can take on positive and negative values, where the sign reveals the orientation of the plane of preferential polarization.

In the literature, the polarization of comets is reported using one of these two definitions. For instance, in application to the case of 21P/G-Z, Kurchakov et al. (1986) and Kiselev et al. (2000) reported results of their polarimetric observations solely using Eq. (2); whereas, Martel et al. (1960) studied polarization only in the sense of Eq. (4). However, as demonstrated in Chernova et al. (1993) and in Table 1 of this study, at large phase angles, there is very little difference between the values of $P_r$ and $P$. Therefore, in current work we analyze and compare the results of our own measurements using Eq. (2) with other observations regardless of whether they consider $P_r$ or $P$.

*2.2 Photometry*

Photometric observations of comet 21P/G-Z were carried out at the observation station Lisnyky of the Astronomical Observatory of Taras Shevchenko National University of Kyiv (Ukraine) with the AZT-8 telescope ($D$ = 0.7 m, F = 2.8m, observatory code – 585) on September 16, 2018. The CCD PL47-10 FLI camera with 1 × 1 binning is used as a detector with an image size of 1027 × 1056 pixels and a scale of 0.95 ″/pix which corresponds to a full field of 16′ 15″ × 16′ 43″. We obtain images through the Johnson-Cousins broadband $V$ filter ($\lambda$ = 0.551 μm, FWHM = 0.088 μm) and $R$ filter ($\lambda$ = 0.658 μm, FWHM = 0.138 μm). A detailed log of observations of comet 21P/G-Z is presented in Table Y.

The reduction procedure, including bias subtraction, dark and flat field corrections, and cleaning cosmic-ray tracks, was carried out using standard methods of photometry using the ITT IDL routines. The morning sky obtained through each filter was used to provide a flat-field correction. To find the instrumental magnitude, the sky background was subtracted. The field stars were used as standards stars. The stellar magnitudes of the standard stars were taken from the NOMAD catalogue (I/297/out in the Vizier database). This catalogue combines the Hipparcos, Tycho-2, UCAC2, USNO-B1.0 et 2MASS catalogues. The seeing value measured as the FWHM of the stars on images was an average of 3.7 arcsec during our observations.

The apparent magnitude of the comet in each filter was determined within the aperture centered at the expected location of its nucleus and with a projected diameter of approximately 6,100 km. The position of the comet nucleus was estimated using the central contour of isophotes closest to the maximum of the comet brightness. To determine its magnitude, the zenith distance was also taken into account. The nightly magnitude of the comet in each filter was averaged over that night's observations. The maximum uncertainty in the magnitude measured with the $V$ filter was as high as 0.06 m.

In the next step, we infer the color of the comet from its magnitude in the $V$ and $R$ filters, taking into account the color of the Sun in the same filters. We notice that the color index of the Solar radiation with those filters previously was reported to be 0.368 (Burlov-Vasiljev et al. 1995; 1998). Therefore, the color index of the innermost 6,000-km coma of 21P/G-Z appears to be 0.13 ± 0.08. It is important that we also try to infer the color index in other places in the 21P/G-Z coma, however, they appear at very large uncertainty, making them meaningless for further analysis. Therefore, in what follows we consider data obtained solely with the 6,000-km aperture centered at the 21P/G-Z nucleus.

In the literature, color of comets is often characterized with the *color slope* that is defined as follows (e.g., A'Hearn et al. 1984; Lamy et al. 2011):

$$S' = \frac{10^{0.4\Delta m} - 1}{10^{0.4\Delta m} + 1} \times \frac{20}{\lambda_2 - \lambda_1} \text{ [\% per 0.1 μm].} \qquad (5)$$

Here $\Delta m$ stands for the color index of the comet reduced for the color index of the Sun. We found it to be $\Delta m$ = 0.13 ± 0.08 in the 6,100-km aperture. Effective wavelengths of the used filters $\lambda_2$ and $\lambda_1$ are measured in μm ($\lambda_2 > \lambda_1$). Thus, the color slope $S'$ is normalized to the wavelength difference and, therefore, it is independent of a specific choice of the filters. Using

Eq. (5), we infer the color slope $S'$ = (11.73 ± 7.75)% per 0.1 μm in the 6,100-km inner coma in 21P/G-Z.

## 3. Results and discussion

The four panels in Fig. 2 show polarization maps of comet 21P/G-Z acquired on September 10, 12, 16, and 17 of 2018. What immediately emerges from this figure is that the measured polarization experiences significant temporal and spatial variations. We address these variations in detail below. Before we do this, we first consider the aperture-averaged polarimetric response in order to make a comparative analysis of data obtained in the 2018 apparition with the former apparitions of comet 21P/G-Z. From the data shown in Fig. 2, we derive the degree of linear polarization within a 6,100-km circular diaphragm centered at the photometric maximum that nearly coincides with the location of the 21P/G-Z nucleus. These results are shown in Fig. 3 with open green circles, along with polarimetric data obtained in the previous apparitions obtained in the blue-green part of spectrum, including the data adapted from Kurchakov et al. (1986). Blue and green diamonds here correspond to the polarization measured with the wideband *B* and *V* filters, and blue squares correspond to two observations conducted using the *IHW blue continuum* (*BC*) filter. As one can see, there is a great deal of overlap between the data obtained with the *B* filter and with the *IHW BC* filter, revealing a very weak gaseous contamination of signal measured with the wideband filters. It is worth noting that the polarimetric observations reported by Kurchakov et al. (1986) were made using apertures ranging from 5500 km to 5900 km that appears in accordance with the 6,100-km aperture used by us in Fig. 3. The observations by Kurchakov et al. could suggest some short-term variations of the degree of linear polarization. Large uncertainty in their measurements, however, do not allow this conclusion with confidence. Our measurements are accompanied with much smaller error bars and suggest variations much greater than the measurement uncertainty. For instance, the rapid decrease of polarization *P* between September 16 and 17 of 2018 is nearly an order of magnitude greater than the measurement uncertainty, dropping ~2.3%, from $P$ = (21.93 ± 0.26)% to $P$ = (19.66 ± 0.34)%.

We note that Kurchakov et al. (1986) also found a similar decrease of polarization measured using the *R* filter between September 19 and 22 of 1985, when the degree of linear polarization decreased from $P$ = (19.1 ± 0.4)% to $P$ = (14.6 ± 1.4)%, which is a drop of more than double the measurement uncertainty. Unfortunately, no *V* filter was used on the latter epoch. Nevertheless, this appears qualitatively consistent with our findings.

Fig. 3 also shows three values of the degree of linear polarization measured in comet 21P/G-Z with the *IHW BC* filter by Chernova et al. (1993) and two values obtained with a non-standard blue continuum filter (λ = 0.443 μm) in Kiselev et al. (2000). In both cases, the polarimetric response is averaged over a considerably larger aperture, 15,000+ km. Finally, we also reproduce no-filter polarimetric observations of the innermost coma of 21P/G-Z (5500–6000 km) that were reported by Martel (1960). In this later case, no error bars were provided.

As one can see in Fig. 3, all the data tend to group along a single curve that, coincidentally, resembles the angular profile of polarization of comet C/1996 B2 (Hyakutake) using the *IHW BC* filter (as reported by Kikuchi 2006). These observational data are shown in Fig. 3 with black points. In addition, we demonstrate their fit (black solid line) that was obtained with a two-component model inferred by Zubko et al. (2016). Although this particular modeling result cannot be extrapolated for the case of 21P/G-Z, because comet C/1996 B2 (Hyakutake) had a persistent red polarimetric color that does not hold for 21P/G-Z, the two-component framework has proven capable of reproducing the photometric and polarimetric observations of numerous comets (e.g., Zubko et al. 2014; 2015; 2016; Ivanova et al. 2017; Picazzio et al. 2019; Luk'yanyk et al. 2019) and are consistent with *in situ* findings (e.g., Zubko et al. 2012).

To model the shape of the micron-sized cometary dust particles we use the *agglomerated debris particles* (cf., Zubko et al., 2012). These particles have highly disordered and fluffy morphology (packing density of constituent material ~0.236) that resembles the shapes of cometary dust particles on the same scale (see, Fig. 4). The light scattering from these particles were compared with laboratory optical measurements of various cometary-dust analogs. It was demonstrated that their light-scattering properties closely matched when the size distributions and complex refractive index $m$ were set to the same values as the measured samples, thus demonstrating that such model particles could be used to obtain accurate retrievals of microphysical properties of the samples (e.g., Zubko 2015; Videen et al. 2018). Such a test significantly raises reliance of retrievals of microphysical properties of dust particles inferred from the astronomical observations of comets.

The two-component model assumes that the cometary coma is populated primarily by two types of dust. One consists of a weakly absorbing material (mainly Im($m$) ≤ 0.02) and the other, a highly absorbing material (Im($m$) ≥ 0.4). The former constraint on the imaginary part of refractive index appears consistent with Mg-rich silicates (Dorschner et al. 1995), and the latter one with carbonaceous materials, such as organics and amorphous carbon (e.g., Duley 1984; Jenniskens 1993). Both types of materials are long known to be abundant species of cometary dust (e.g., Fomenkova et al. 1992; Ishii et al. 2008). This model may reproduce the vast majority of polarimetric observations of comets by fitting only a single free parameter, the volume ratio of the weakly absorbing component to the highly absorbing component (Zubko et al. 2016). It is worth noting that the two-component model can reproduce the phase function and photometric color of dust in comets (Zubko et al. 2014; Ivanova et al. 2017; Picazzio et al. 2019; Luk'yanyk et al. 2019), as well as significant spatial variations of linear polarization observed in some comets (Zubko et al. 2012; 2015). Observations of some rare comets suggest domination of only one type of dust in their coma, at least over short time period (e.g., Picazzio et al. 2019).

Using the two-component approach, we model the polarimetric response in the *V* filter of comet 21P/G-Z. We include in the analysis the five observations by Kurchakov et al. (1986) and four of our own observations. However, we also take into consideration two observations without filters made by Martel et al. (1960). The latter seems reasonable in view of the weak wavelength dependence of the polarization of 21P/G-Z in the visible. Thus, 11 data points embrace the phase-angle range from α = 71.1° to 87.5°. We emphasize that there are numerous options for fitting these data within the two-component model. Therefore, we place further constraints on modeling by incorporating polarimetric observations made using the wideband *R* filters, even though most of them were not synchronized with the observations made using the *V* filter and often correspond to a much larger aperture. The *R*-filter data have been adapted from Kurchakov et al. (1986), Chernova et al. (1993), and Kiselev et al. (2000). However, because there is a mismatch in circumstances of observations with the *V* and *R* filter, we are not searching for the best fit to polarization in red light.

In the top panel of Fig. 5, we present modeling results for polarization in the *V* filter that are labeled as *Fit #1* and *Fit #2*. In the modeling fits throughout this paper, we use the same refractive index of highly absorbing particles $m = 2.43 + 0.59i$, which corresponds to amorphous carbon in the visible (Duley 1984). However, we investigate different options for weakly absorbing particles. For instance, within Fit #1, we consider refractive index $m = 1.6 + 0.0i$, and in Fit #2 we consider $m = 1.6 + 0.02i$. Such refractive indices are representative of Mg-rich silicates with small Fe content (Dorschner et al. 1995). We remove one free parameter by assuming the weakly and highly absorbing particles obey the same power-law size distribution $r^{-n}$ with the power index $n = 1.7$. It is worth noting that at refractive indices selected for weakly and highly absorbing particles in Fig. 5, a reasonably good fit is possible in the range of power index $n = 1.7 \pm 0.2$. These appear in good quantitative accordance with *in situ* findings for micron-sized dust particles in comet 1P/Halley (e.g., Mazets et al. 1986) as well as with models of other comets (e.g., Zubko et al. 2016). The observed power index $n$ clearly tends to increase with particle size, and, in particles having size in excess of 10 μm, it is $n \geq 3$ (Mazets et al. 1986; Price et al. 2010). Further details of this modeling are discussed by Zubko et al. (2016).

In the top panel of Fig. 5, Fits #1 and #2 reproduce the angular profile of polarization in comet 21P/G-Z reasonably well. These fits were performed from combined polarimetric data obtained during almost 60 years (i.e., 9 full revolutions around the Sun), suggesting not much change in cometary constituents over this time period; at least, with regard to dust composition. We note that Fit #1 is obtained with a relative volume of weakly absorbing particles being of 23% and, correspondingly, 77% of highly absorbing particles. In Fit #2, the relative abundance of weakly absorbing particles is somewhat greater, 32%. Thus, an increase of Im($m$) in the weakly absorbing component tends to increase its relative abundance, which is consistent with the modeling results of Zubko et al. (2016).

We consider the wavelength-independent refractive indices in Fits #1 and #2 as the same holds for many other comets (e.g., Zubko et al. 2014; 2016). As one can see in the bottom panel of Fig. 5, such particles yield polarization slightly higher in the *R* filter compared to the *V* filter at phase angle ranging from $\alpha \approx 40°$ up to 80–90°; although their maximum linear polarization $P_{max}$ is somewhat smaller at longer wavelength. Therefore, Fits #1 and #2 may reproduce four out of six observations by Kurchakov et al. (1986) with red polarimetric color. However, modeling the other observations in the *R* filter require further model refinement. This can be accomplished in practice by introducing a wavelength dependence of the imaginary part of refractive index Im($m$). Therefore, in the *V* filter we set Im($m$) = 0.02 (the same as in Fit #2) and in the *R* filter we set Im($m$) = 0.01 (as in Fit #1). This extended model is labeled with *Fit #2/1*. What is significant about this type of wavelength dependence of Im($m$) is that it is consistent with trends in natural materials in the visible and NIR where absorptivity is greater at shorter wavelength (e.g., Dorschner et al. 1995; Jenniskens 1993). The volume ratio of the weakly absorbing particles and highly absorbing particles in Fit #2/1 is the same as in Fit #2. As one can see in the bottom panel of Fig. 5, Fit #2/1 reveals unambiguously blue polarimetric color and, therefore, is consistent with the other four polarimetric observations of 21P/G-Z in the *R* filter. Within this approach, variations in polarimetric color of comet 21P/G-Z could be attributed to a slight change in chemical composition of its dust population, e.g., a slightly different iron content of the Mg-rich silicates (Dorschner et al. 1995) or some contamination of Mg-rich silicates with organics that have a strong wavelength dependence of Im($m$) (Jenniskens 1993).

Finally, we address the polarization measured in comet 21P/G-Z with the *V* filter on 2018-09-17. Fits #1 and #2 match all polarimetric observations presented in the top panel of Fig. 5, except for this particular epoch. This observation may reflect a somewhat different balance between weakly absorbing particles and highly absorbing particles in the 21P/G-Z coma on 2018-09-17. Indeed, Fit #2 produced using a volume ratio of 32% weakly absorbing particles can be adjusted to a lower polarization on 2018-09-17 by considering a ratio of 41.5% weakly absorbing particles. Simultaneously, the adjusted Fit #2/1 mimics the very low *P* in the *R* filter reported by Kurchakov et al. (1986) on 1985-09-22. Thus, a temporal increase of abundance of weakly absorbing particles in the coma that is accompanied with a decrease of positive polarization is seemingly a systematic phenomenon for comet 21P/G-Z that occurs in its different apparitions.

In Fig. 6 we repeat the analysis for Re($m$) = 1.7. In this case, the power index takes on a value $n$ = 2.0, and its possible range being $n$ = 2.0 ± 0.3. Fit #3 is obtained with 22% weakly absorbing particles with Im($m$) = 0.01; whereas, Fit #4 is obtained with 30% weakly absorbing particles with Im($m$) = 0.02. Fit #4/3 can be adjusted to the polarization on 2018-09-17 by considering 38% weakly absorbing particles. Evidently, the relative volumes of the weakly and highly absorbing particles in Fig. 6 nearly coincide with those obtained in Fig. 5. This implies that our retrievals are hardly affected by the assumption of the real part of refractive index Re($m$) of the weakly absorbing particles, at least within the range Re($m$) = 1.6 – 1.7.

To this point, we have only considered the phase dependence of the degree of linear polarization using different filters. However, one can also compute reflectivity of dust particles in those filters and, thus, characterize the color slope $S'$ (Eq. 5). We perform such computations at phase angle $\alpha$ = 77° that corresponds to the photometric observations made on 2018-09-16. We note that color is dependent on the geometry of observations and, hence, on phase angle $\alpha$ (e.g., Zubko et al. 2014). The model yields $S'$ = 2.53% per 0.1 µm in Fit #1, $S'$ = 3.75% per 0.1 µm in Fit #2, $S'$ = 9.42% per 0.1 µm in Fit #2/1, $S'$ = 2.41% per 0.1 µm in Fit #3, $S'$ = 3.49% per 0.1 µm in Fit #4, and $S'$ = 8.67% per 0.1 µm in Fit #4/3. As one can see, Fits #1, #2, #3, and #4 are hardly consistent with the color slope inferred from the photometric observations with the 6,100-km aperture on 2018-09-16 $S'$ = (11.73 ± 7.75)% per 0.1 µm. However, Fits #2/1 and #4/3 appear in accordance with the photometric observations within their error bars. Both these fits suggest blue polarimetric color in comet 21P/G-Z on 2018-09-16.

We now consider the spatial and temporal variations of the degree of linear polarization shown in Fig. 2. What immediately emerges from the polarization maps are significant day-to-day variations of polarization in the 21P/G-Z coma. For instance, on 2018-09-16 in the tailward direction (toward the top in Fig. 2), there is an extended dust cloud with low polarization. This feature was very faint four days earlier, on 2018-09-12, and weakens again only one day later, on 2018-09-17. Within this epoch, the comet was at a relatively small heliocentric distance, $r_h$ ≈ 1 au, where water ice sublimates very quickly. The lifetime of even 10-µm particles consisting of pure water ice is only few hours; whereas, dirty-ice particles would totally sublimate within only one-two minutes (Beer et al. 2006). Therefore, such dust particles located at 1,000+ km from the 21P/G-Z nucleus cannot be expected to be transformed by water-ice sublimation. Instead, the disappearance of the dust cloud with low polarization could occur due to particle dynamics.

In order for dust particles to leave the field of view within a day, they should move at high velocities. Their velocity component projected on the image of the comet in Fig. 2 should be as high as ~700 m/s; whereas, its absolute value could exceed 1 km/s. Such speeds cannot be acquired from cometary gases, whose expansion velocity varies with the heliocentric distance $r_h$ as follows: $V_{gas} = \dfrac{0.85}{\sqrt{r_h}}$ km/s (Combi et al. 2004). Evidently, at $r_h$ ≈ 1 au, dust particles move faster than the gas molecules.

Such high velocities can be acquired from solar-radiation pressure (e.g., Fulle 2004). Its effect is often characterized with the parameter β that is the ratio of the radiation-pressure force over the solar-gravity force. As demonstrated in Zubko et al. (2016), low positive linear polarization in comets is produced by presence of weakly absorbing, Mg-rich silicate particles. Presence of such particles in 21P/G-Z is suggested above. On the other hand, in micron-sized particles (i.e., $r$ = 0.5 µm), having Mg-rich silicate composition, the β parameter varies from

0.538 to 1.284, depending on their Fe content (Zubko et al. 2015). This allows us to estimate how long such particles need to gain speed of 700 m/s from the radiation-pressure effect. In the assumption of zero initial velocity and unit heliocentric distance, non-absorbing particles with $\beta = 0.538$ are accelerated to ~700 m/s within 2.54 days; whereas, slightly absorbing particles with $\beta = 1.284$ acquire such speeds in only 1.06 days. Under constant pressure, velocity is proportional to time. Thus, we can conclude that the dust cloud with low linear polarization observed on 2018-09-16 was released from the nucleus within a few days of observation and, it consist of Mg-rich silicate particles. Any dust released more than a few days prior would have achieved speeds from the solar-radiation pressure that would have driven them out of the field of view. The approximately one-day time period of replenishment is illustrated in Fig. 2, which shows a greatly reduced cloud on 2018-09-17, only a day after the cloud had much greater optical depth.

The state of the coma is in constant flux, with particles entering from the nucleus and leaving due to radiation pressure. However, the radiation pressure is not constant for all particles. Micron-sized carbonaceous particles have systematically greater values of the $\beta$ parameter than pure silicates. For instance, in micron-sized amorphous-carbon particles, $\beta \approx 2.13$, compared with $\beta \approx 0.54 – 1.28$ in silicate particles (Zubko et al. 2015). This results in the amorphous-carbon particles being accelerated much more efficiently. For instance, while the silicate particles with $\beta \approx 1.284$ accelerate to ~700 m/s in approximately one day, the amorphous-carbon particles accelerate to ~1100 m/s. Therefore, the amorphous-carbon particles leave the field of view much more rapidly, which can result in a change of coma composition with time.

Finally, we consider the cloud located in the sub-solar part of the coma, beneath the white cross in Fig. 2, that produces the high polarization. We consider here three zones of integration of the polarimetric response that are marked with black circles. Although the linear polarization of light scattered from this cloud noticeably varies from date to date, the feature remains overall intact through the entire period of observations. This would be impossible without regular replenishment of dust particles forming the cloud; whereas, variations of linear polarization could reflect the local and short-term changes in the volume ratio of the silicate and amorphous-carbon particles. For example, in Table 3, we present a relative volume concentration of the silicate particles inferred by adjusting Fits 2 and 2/1 to the polarimetric responses in the 6,100-km circular aperture centered at the 21P/G-Z nucleus and in the 2,000-km, 5,100-km, and 10,200-km apertures displaced from the nucleus location as shown in Fig. 2. Unlike in Figs. 5 and 6, we do not search for the best fit to the entire set of observational data, but only to specific values of polarization in the given part of coma. Note also that the relative volume concentration of the amorphous-carbon particles is complimentary to the data presented in Table 3.

What we can observe from Table X are significant temporal and spatial variations of the relative abundance of silicate particles. Their low-volume concentration correlates with high positive polarization that is the best embraced with the 2,000-km aperture. The lowest relative abundance of silicate particles, 0.24 ± 0.04, was detected on 2018-09-12. However, the relative volume concentration of silicate particles clearly increases with aperture size; whereas, their largest abundance, 0.451 ± 0.006, was found with the 10,200-km aperture on 2018-09-17. Such a trend can be explained with a twice greater efficiency of solar-radiation pressure that the amorphous-carbon particles experience. Due to the effect of the solar-radiation pressure, these carbonaceous particles get quickly washed out from the inner coma giving an increase of the relative abundance to the silicate particles.

It also is worth noting that a domination of dust particles with high carbon content was previously noticed in cometary jets (Zubko et al. 2012; 2015). Although we do not resolve these features in the images of the innermost coma, the high abundance of carbonaceous particles within the smallest 2,000-km aperture that is placed in close proximity to the nucleus appears to suggest jet activity in comet 21P/G-Z. Simultaneously, day-to-day variation of the linear polarization in the 2,000-km aperture and, thus, relative abundance of the amorphous-carbon particles, suggest the jet activity is discontinuous, occurring with some short-term breaks.

## 4. Conclusion

We report results of our polarimetric observations of comet 21P/G-Z with the wideband $V$ filter at large phase angles, $\alpha \approx 76$–$78°$, between 10 and 17 of September, 2018. Furthermore, on 2018-09-16, simultaneously with polarimetric observations, we also obtained the $V$–$R$ color in the 21P/G-Z innermost coma. Analysis of this entire set of observational data reveals that they can be satisfactorily reproduced within a two-component model of cometary coma that was previously exploited in application to numerous comets. It appears that the 21P/G-Z coma predominantly consists of Mg-rich silicate particles (Re($m$) = 1.6 – 1.7 and Im($m$) = 0.01 – 0.02) and amorphous-carbon particles ($m$ = 2.43 + 0.59$i$). Both components were found to obey the same power-law size distribution with the power index ranging from $n$ = 1.5 to 2.3 Spatial and temporal variations of the polarimetric response in the 21P/G-Z coma, then, are explained with variations of the relative abundance of these two types of particles and their high-speed motion under the solar-radiation pressure, ~1 km/s. Analysis of the 21P/G-Z polarimetric images suggest two types of activity in its coma that happens on irregular and regular basis; whereas, in the latter case, we find indirect evidence for a periodical jet-type activity.

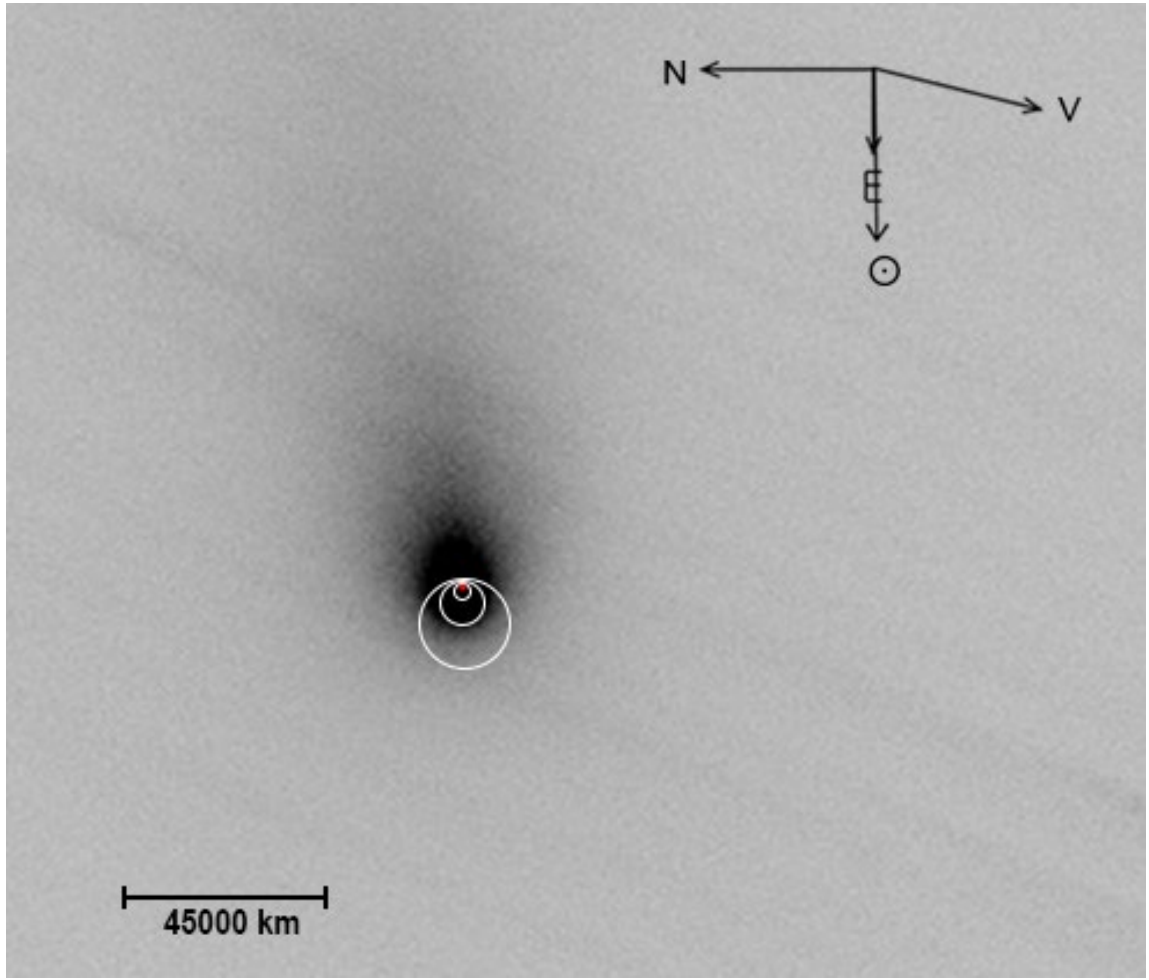

Fig. 1: The resulting image of comet 21P/G-Z taken through the *V* filter at the Ussuriysk Astrophysical Observatory on September 16, 2018, at the phase angle α = 77.8°.

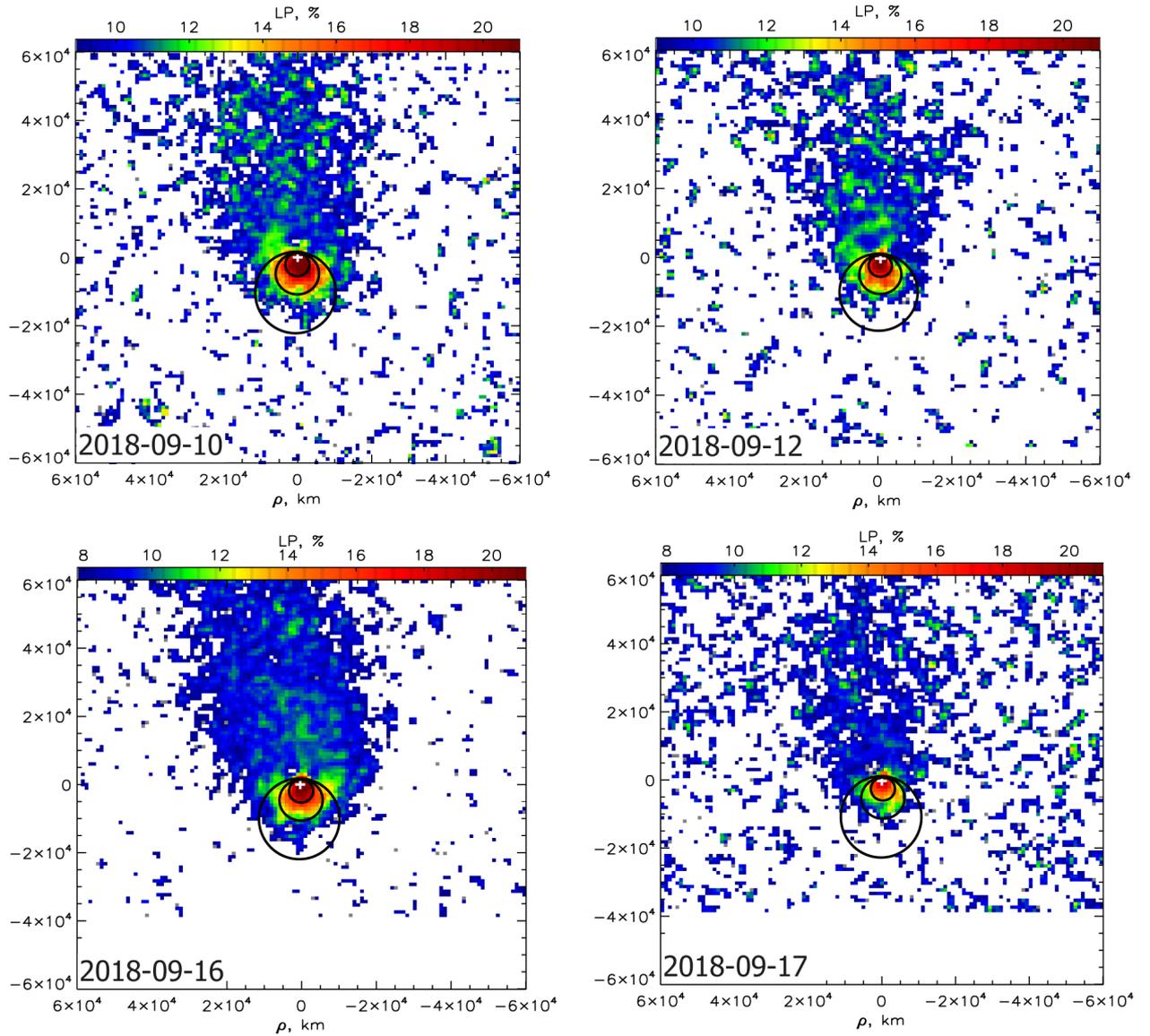

Fig. 2: Polarimetric images of comet 21P/G-Z obtained with the *V* filter on September 10, 12, 16, and 17 of 2018. The Sun is on the bottom, north is on the left, and the white cross denotes the nucleus location.

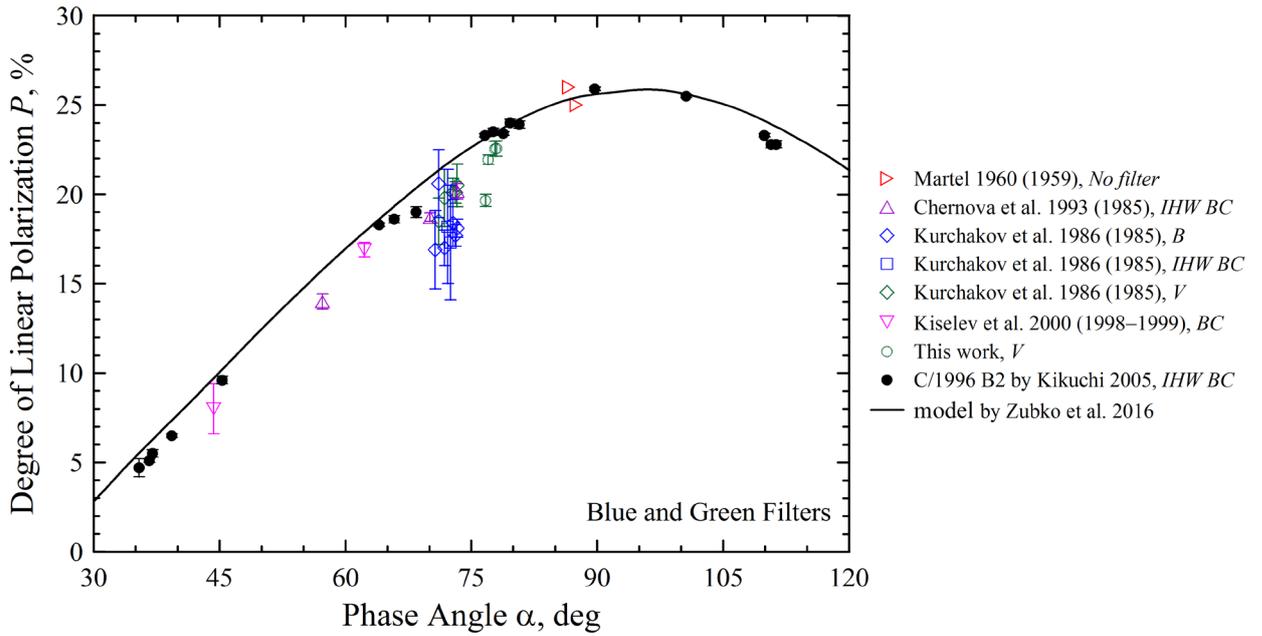

Fig. 3: Aperture-averaged degree of linear polarization $P$ in the blue-green portion of the spectrum as a function of the phase angle $\alpha$. Open symbols show data for comet 21P/G-Z obtained by different teams on different epochs (see legend). Black points present polarization in comet C/1996 B2 (Hyakutake) measured using the *IHW BC* filter (data adapted from Kikuchi 2006), and the black solid line is a fit to those observations obtained with a two-component model of cometary dust (adapted from Zubko et al. 2016).

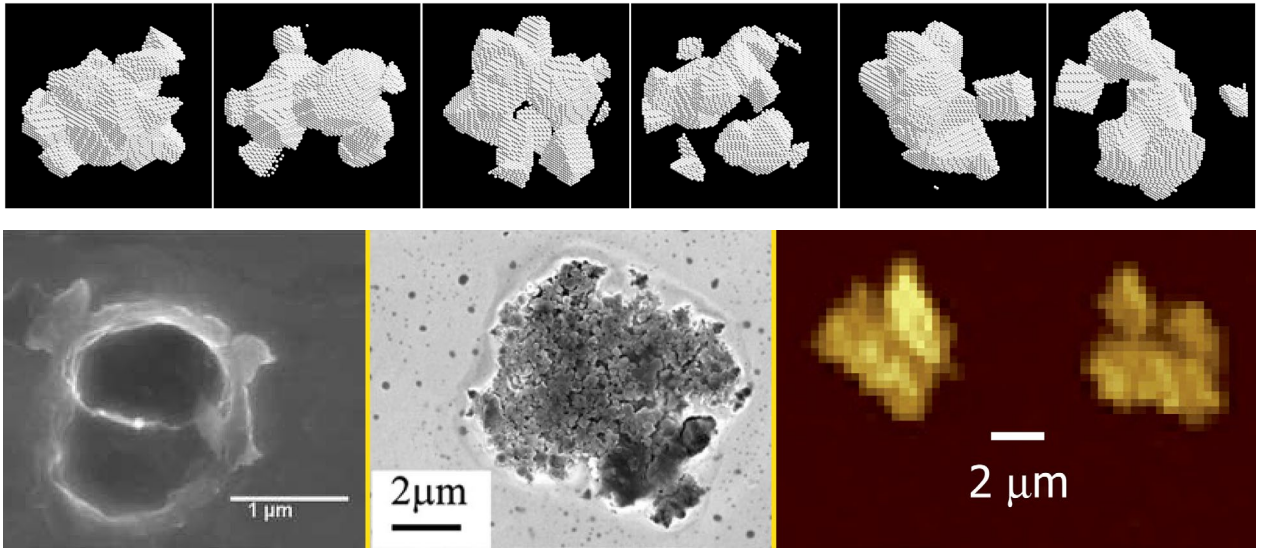

Fig. 4: Top: Images of six samples of irregularly shaped *agglomerated debris particles* that are used in the two-component model of cometary dust. Bottom: Micron-sized crater in the Al foil covering the *Stardust* sampler module that is presumably produced by a dust particle from comet 81P/Wild 2 (left), a cometary dust particle collected in the stratosphere and highly likely originated from comet 26P/Grigg–Skjellerup (middle), and two micron-sized dust particles sampled by *Rosetta* in the innermost coma of 67P/Churyumov–Gerasimenko (right). Images on the bottom are adapted from Hörz et al. (2006), Busemann et al. (2009), and Bentley et al. (2016), respectively.

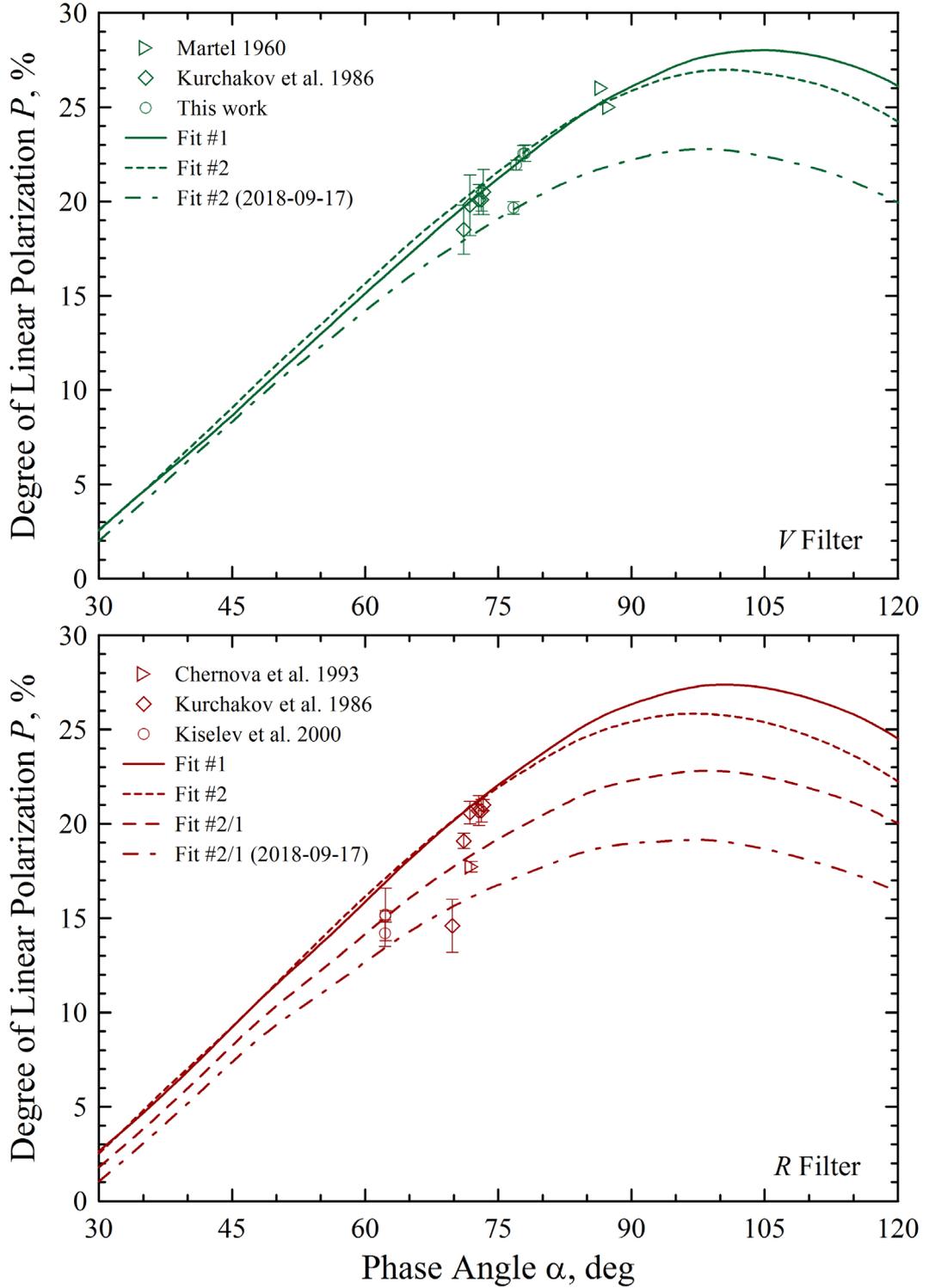

Fig. 5: Modeling of polarization in comet 21P/G-Z in the wideband *V* filter (top) and the wideband *R* filter (bottom). The fit is obtained in the two-component framework. A highly absorbing component having refractive index $m = 2.43 + 0.59i$ corresponding to amorphous carbon, and a weakly absorbing component $Re(m) = 1.6$ and $Im(m) = 0.01 - 0.02$. See text for more details on the modeling parameters.

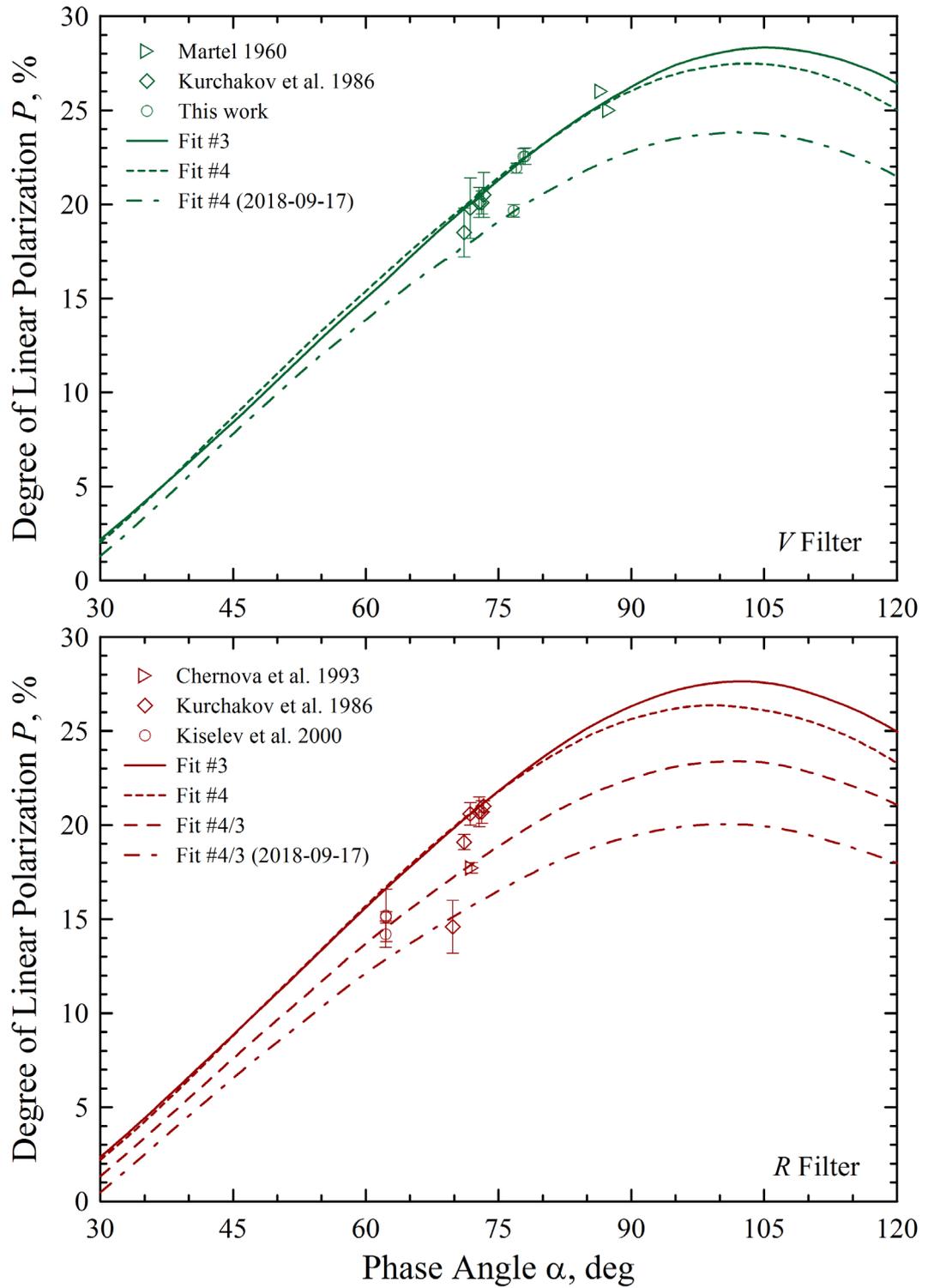

Fig. 6: The same as in Fig. 5, but with Re($m$) = 1.7 for the weakly absorbing component.

**Tables**

Table 1: Log of observations of comet 21P/G-Z in September of 2018.

| Date, UT | Exposure, sec | Filter | Numbers | Airmass | Observatory Code |
|---|---|---|---|---|---|
| 10.7310 | 30 | V | 45 | 1.560 | C15 |
| 12.7267 | 30 | V | 45 | 1.627 | C15 |
| 16.6875 | 60 | V | 45 | 2.524 | C15 |
| 16.9894* | 20 | V, R | 7, 8 | 1.96 | 585 |
| 17.7125 | 30 | V | 45 | 2.150 | C15 |

Table

| Date, UT | r, au | Δ, au | α, ° | P, % | θ, ° | $P_r$, % |
|---|---|---|---|---|---|---|
| **6,109 km** | | | | | | |
| 10.7310 | 1.01 | 0.39 | 78.0 | 22.56±0.43 | 3.13 | 22.93±0.42 |
| 12.7267 | 1.01 | 0.39 | 77.8 | 22.54±0.42 | 2.91 | 22.35±0.42 |
| 16.6875 | 1.01 | 0.39 | 77.0 | 21.93±0.26 | 12.17 | 19.82±0.23 |
| 17.7125 | 1.01 | 0.40 | 76.7 | 19.66±0.34 | 10.75 | 18.16±0.31 |
| **2,036 km** | | | | | | |
| 10.7310 | 1.01 | 0.39 | 78.0 | 22.30±1.3 | 2.02 | 22.24±1.29 |
| 12.7267 | 1.01 | 0.39 | 77.8 | 25.22±1.3 | 1.74 | 25.17±1.29 |
| 16.6875 | 1.01 | 0.39 | 77.0 | 23.07±0.56 | 15.17 | 19.91±0.48 |
| 17.7125 | 1.01 | 0.40 | 76.7 | 23.46±1.3 | 6.89 | 22.78±1.26 |
| **5,091 km** | | | | | | |
| 10.7310 | 1.01 | 0.39 | 78.0 | 21.5±0.6 | 1.67 | 21.20±0.59 |
| 12.7267 | 1.01 | 0.39 | 77.8 | 24.13±0.63 | 4.62 | 23.80±0.62 |
| 16.6875 | 1.01 | 0.39 | 77.0 | 22.54±0.42 | 14.16 | 19.77±0.36 |
| 17.7125 | 1.01 | 0.40 | 76.7 | 19.9±0.44 | 14.11 | 17.53±0.38 |
| **10,183 km** | | | | | | |
| 10.7310 | 1.01 | 0.39 | 78.0 | 21.53±0.21 | 2.83 | 21.42±0.20 |
| 12.7267 | 1.01 | 0.39 | 77.8 | 21.57±0.32 | 3.51 | 21.40±0.31 |
| 16.6875 | 1.01 | 0.39 | 77.0 | 19.23±0.18 | 12.38 | 17.46±0.16 |
| 17.7125 | 1.01 | 0.40 | 76.7 | 18.74±0.15 | 11.20 | 17.32±0.14 |

Table 3. Degree of linear polarization and retrievals of relative volume concentration of the silicate particles obtained within Fits 2 and 2/1 in different parts of comet 21P/G-Z.

| Date, UT | 6,109 km | | 2,036 km | | 5,091 km | | 10,183 km | |
|---|---|---|---|---|---|---|---|---|
| | $P$, % | Rel. Vol. | $P$, % | Rel. Vol. | $P$, % | Rel. Vol. | $P$, % | Rel. Vol. |
| 10.7310 | 22.56 ± 0.43 | 0.323 ± 0.014 | 22.30 ± 1.30 | 0.333 ± 0.043 | 21.50 ± 0.60 | 0.359 ± 0.020 | 21.53 ± 0.21 | 0.357 ± 0.007 |
| 12.7267 | 22.54 ± 0.42 | 0.324 ± 0.014 | 25.22 ± 1.30 | 0.244 ± 0.036 | 24.13 ± 0.63 | 0.274 ± 0.018 | 21.57 ± 0.32 | 0.356 ± 0.011 |
| 16.6875 | 21.93 ± 0.26 | 0.332 ± 0.009 | 23.07 ± 0.56 | 0.295 ± 0.018 | 22.54 ± 0.42 | 0.312 ± 0.014 | 19.23 ± 0.18 | 0.431 ± 0.008 |
| 17.7125 | 19.66 ± 0.34 | 0.414 ± 0.013 | 23.46 ± 1.30 | 0.284 ± 0.040 | 19.90 ± 0.44 | 0.405 ± 0.016 | 18.74 ± 0.15 | 0.451 ± 0.006 |